\documentclass[11pt, a4paper]{article}
\usepackage{jcappub}
\usepackage{natbib}
\usepackage{aas_macros}

\usepackage[T1]{fontenc}
\usepackage[utf8]{inputenc}
\usepackage[english]{babel}
\usepackage{amsmath}
\usepackage{amssymb}
\usepackage{bm}

\usepackage{graphicx}
\usepackage{subfig}
\usepackage{enumerate}
\usepackage{epsfig,epstopdf}
\usepackage{color,xcolor}
\usepackage{multirow}
\usepackage{mleftright}
\usepackage{soul}

\usepackage{url}
\usepackage{xspace}
\usepackage{comment}
\usepackage{bm}
\usepackage{braket}
\usepackage[noabbrev]{cleveref} %noabbrev,nameinlink
\usepackage{hhline}

\usepackage{hyperref}
\long\def\comment#1{}

\usepackage{float}

\usepackage{tikz} 
\usetikzlibrary{arrows,patterns} 
\definecolor{myblue}{RGB}{0,0,255} 
\definecolor{myred}{RGB}{255,0,0}
\usepackage{subcaption}

\graphicspath{{./plots/}}
\graphicspath{{./plots/}}

\title{{Exploring Beyond $\Lambda$CDM with the Weak Lensing Power Spectrum and Bispectrum}}

\author[1,2]{Liantsoa F. Randrianjanahary}
\author[1]{Chandrachud B. V. Dash}
\affiliation[1]{Astrophysics Research Centre $\&$ School of Mathematics, Statistics and Computer Science, University of KwaZulu-Natal, Durban, 4041, South Africa}
\affiliation[2]{Department of Physics $\&$ Astronomy, University of the Western Cape, Cape Town 7535, South Africa}

\emailAdd{fina.liantsoarandrianjanahary@gmail.com} \emailAdd{cb.vaswar@gmail.com}

\abstract{We forecast how well the tomographic weak-lensing power spectrum and
bispectrum measured by a Stage~IV survey can constrain three physically
distinct extensions of $\Lambda$CDM: the CPL parametrization of dynamical dark
energy, an interacting dark energy (IDE) model with energy--momentum exchange
in the dark sector, and Hu--Sawicki $f(R)$ gravity. Within the Fisher
formalism, and marginalising over the cosmological parameters together with the
photometric-redshift scatter $\sigma_z$ and the intrinsic-alignment amplitude
$A_{\rm IA}$, we compare the power spectrum alone with the combined power
spectrum and bispectrum. Adding the bispectrum tightens the constraints most
for the CPL and $f(R)$ models: the marginalised $1\sigma$ error on the CPL
equation of state improves from $\sigma(w_0)=0.085$ to $0.061$, and on the
Hu--Sawicki amplitude from $\sigma(\log_{10}|f_{R0}|)=0.198$ to $0.129$,
corresponding to gains in the two-parameter figure of merit of $1.74$ and
$1.64$ respectively. For interacting dark energy the improvement is more
modest, $\sigma(\alpha)=0.0123$ to $0.0112$ (a figure-of-merit gain of $1.12$),
because in our treatment the coupling enters the nonlinear spectrum mainly
through a rescaling of the growth. On its own the bispectrum is always the
weaker probe, but in combination with the power spectrum it breaks parameter
degeneracies and improves every forecast. These results show that the
weak-lensing bispectrum is a practical way to increase the scientific return of
Stage~IV surveys, with a benefit that depends on the model.}

\begin{document}

\maketitle
\clearpage
\section{Introduction}
\label{intro}
Understanding the physical origin of the late-time cosmic accelerated expansion is one of the key challenges in contemporary cosmology \citep{2018beyo.book.....B, Albrecht:2006um,huterer2017dark, Bull:2015stt}. Several independent observations, including the cosmic microwave background (CMB), baryonic acoustic oscillations (BAO), redshift space distortions, type Ia supernovae, and cosmic chronometers (CC), support the flat $\Lambda$CDM, where the cosmological constant drives cosmic acceleration \citep{Guzzo:2008ac,Carroll:1991mt,SupernovaCosmologyProject:1998vns,SupernovaSearchTeam:1998fmf}. Although $\Lambda$CDM provides an excellent phenomenological description of current observational data, it is plagued by several fundamental theoretical shortcomings, e.g., fine-tuning and coincidence problems.  This motivates to search for alternative frameworks that model independently reconstruct the dark energy equation of state (EoS) or modify General Relativity (GR) \citep{perivolaropoulos2022challenges}. Also recently DESI’s growing evidence for dynamical dark energy ($2.5$–$3.9\sigma$ preference for dark energy equation of state present day value $w_0 > -1$ and time derivative $w_a < 0$; \citep{Adame_2025}) demands independent confirmation from complementary probes and hunting for alternatives of $\Lambda$. In the literature, authors explore many such myriads of phenomenologically well motivated models beyond $\Lambda$CDM \citep{perivolaropoulos2022challenges}. Discriminating among these competing scenarios demands observables that are simultaneously sensitive to both the background expansion history and the growth of cosmic structure. Weak gravitational lensing stands out as one of the most powerful probes in this regard and has been explored as a prime cosmological probe to exploit beyond $\Lambda$CDM framework \citep{Frugte:2025afv,Benetti:2024dob,huterer2010weak,Dash_2021,dash2022probing,Dinda:2018eyt,Dinda:2017swh}. Because lensing responds directly to the total projected matter distribution, it circumvents uncertainties associated with galaxy bias, making it a clean tracer of large-scale structure. Current generation (Stage-III) surveys like KiDS \citep{Yoon:2025eem}, DES \citep{descollaboration2026darkenergysurveyyear}, Hyper Suprime-Cam \citep{aihara2018hyper} have already placed competitive tomographic constraints on the $\Lambda$CDM parameter space, particularly on the $S_8 \equiv \sigma_8\sqrt{\Omega_m/0.3}$ combination. The forthcoming (Stage IV) program Vera C.\ Rubin Observatory Legacy Survey of Space and Time (LSST) \citep{2009arXiv0912.0201L,lsst2012large}, Euclid \citep{Euclid:2019clj,Euclid:2023wdq}, Nancy Grace Roman Space Telescope \citep{committee2025romanobservationstimeallocation} will extend these measurements to substantially higher source densities and redshifts, and are projected to yield percent-level constraints on both the Hubble expansion rate $H(a)$ and the linear growth factor $D(a;k)$ through measurements of the nonlinear matter power spectrum \citep{Prat:2025ucy,Ishak:2019aay,2009arXiv0912.0201L,lsst2012large}.

Most analyses of weak gravitational lensing involve the study of two-point statistics, including the shear or convergence power spectrum. This signal is highly informative but only describes the Gaussian contribution to the matter density field. Hence, nonlinear gravitational evolution leads to the appearance of non-Gaussian features in the matter density field, as described by the contribution of higher correlation functions. In particular, the weak lensing bispectrum is the harmonic transform of the three-point correlation function and is responsible for mode coupling and nonlinear structure formation. A joint analysis of the power spectrum and the bispectrum has the potential to improve constraints on cosmological parameters significantly \citep{Randrianjanahary:2023rgp,Karagiannis:2022ylq}. This is because it can help to break the degeneracy partially between parameters and can also account for scale-dependent effects that are relevant to the cosmological models under consideration, including the effects of dark energy and modified gravity on the growth of structure. In this paper, we propose a general weak lensing analysis that includes the analysis of convergence power spectra and bispectra. We test a range of cosmological models, including those relevant to dark energy and modified gravity theories. Specifically, we consider the following extensions to the standard $\Lambda$CDM model: (i) $w$CDM with dynamical dark energy; (ii) an interacting dark energy model in which energy and momentum transfer are allowed within the dark sector; and (iii) the Hu-Sawicki $f(R)$ model, in which the Poisson equation is modified and a scale- and time-dependent modification of gravity occurs.

Among a wide range of systematics that can affect the weak lensing signal, two of them are critical that draw our attention in this study, including photometric redshift uncertainty \citep{2006ApJ...636...21M,Bernstein:2009bq} and intrinsic alignment of galaxies \citep{Troxel_2015,Bridle2007,Kirk_2012,Joachimi:2015mma}. Unlike spectroscopic redshift, which is highly accurate, photometry uses broadband colors. Because different galaxy types at different redshifts can produce similar colors, the estimation is inherently uncertain. Photometric redshift uncertainty is the margin of error in estimating a galaxy's distance by analyzing its light through broad-band filters rather than precise spectroscopy. It represents the difference between the estimated and true redshifts. This redshift uncertainty can degrade the cosmological information \citep{Yuan:2019ofk,Bernstein:2009bq,Zhang:2025rxp,Ma:2005rc}. Galaxies are not randomly oriented but tend to align with the large-scale structure in which they reside. This creates a non-lensing correlation that can bias cosmic shear measurements.
Intrinsic alignments can significantly suppress the cosmic shear power spectrum, as demonstrated by \citep{Hirata:2004gc}, and were detected for the first time in the KiDS-450 photometric survey by \citep{Yao:2019eyx, Pedersen:2019wfp}. For weak lensing, one of the main difficulties in making cosmological forecasts was the ability to accurately model the intrinsic alignments of galaxies, i.e., a local orientation of galaxies that acts to mimic the cosmological lensing signal \citep{Euclid:2019clj}. Several physical models have been proposed \citep{Joachimi:2015mma,Kirk:2015nma, Kiessling:2015sma, Lamman:2023hsj, Kacprzak:2022pww, Yoon:2025eem} that model intrinsic alignments in a more realistic manner, drawing increased attention in this area \citep{Euclid2020}.

Our main objective is to quantify the additional cosmological information gained by combining two- and three-point weak-lensing statistics. In particular, we investigate how adding the bispectrum improves constraints on the dynamical equation-of-state parameters, the dark-sector interaction strength, and the $f(R)$ gravity amplitude, relative to a power spectrum-only analysis. This joint approach is especially timely for next-generation surveys, whose statistical precision demands that all available information be extracted in order to test the nature of dark energy and gravity robustly.

The paper is organized as follows. Section~\ref{sec:de_models_background_growth}
introduces the beyond-$\Lambda$CDM models and their background expansion and linear growth. Section~\ref{sec:wl_observables_systematics} describes the weak-lensing observables and the main observational systematics, and Section~\ref{sec:bispectrum} develops the weak-lensing bispectrum formalism, including the Fisher matrix used for the forecasts. Section~\ref{sec:results} presents the analysis and discussion of our results, and we summarize our
conclusions in Section~\ref{sec:conclusions}.

\section[Dark energy models: background expansion and growth]%
        {Dark energy models beyond $\Lambda$CDM: background expansion and growth}
\label{sec:de_models_background_growth}

In this section, we present the standard concordance dark energy model and beyond-$\Lambda$CDM scenarios under study, within a unified framework for background expansion and linear structure growth before discussing the weak lensing observables.
\subsection{Unified description of the background expansion}
\label{subsec:background_expansion}
The homogeneous and isotropic background geometry of the Universe is governed by the Friedmann equations, which relate the Hubble expansion rate $H(a) \equiv \dot{a}/a$ to the energy content of each cosmological fluid \citep{2020moco.book.....D,Dodelson:2003ft,Peebles:2002gy}. In the spatially flat Friedmann-Lemaitre–Robertson–Walker (FLRW) metric, the first Friedmann equation reads as a function of stretch factor $a$
\begin{equation}
  E(a) \equiv \frac{H(a)}{H_0}
  = \left[
      \Omega_{m0}\,a^{-3}
    + (1-\Omega_{m0})\,\mathcal{F}_X(a)
    \right]^{1/2},
  \label{eq:Ea_unified}
\end{equation}
where dimensionless Hubble parameter $E(a) \equiv H(a)/H_0$ and $\Omega_{m0}$ is the matter density at the present time. The term $X\in\{\Lambda{\rm CDM},w{\rm CDM},{\rm IDE},f(R)\}$ labels the model, and $\mathcal{F}_X(a)$ encodes the effective dark energy or modified gravity contribution to the expansion. In the following we specify $\mathcal{F}_X(a)$ for each case.

\subsubsection*{$\Lambda$CDM}
In the concordance model, dark energy is a cosmological constant with equation of state $w=-1$ corresponding to a time-independent vacuum energy density $\rho_\Lambda = \Lambda/(8\pi G)$ \citep{Peebles:2002gy,Condon:2018eqx,weinberg1989cosmological}. Since $\rho_\Lambda$ does not evolve with the scale factor, the modification function is trivially constant,
\begin{equation}
  \mathcal{F}_{\Lambda{\rm CDM}}(a) = 1,
  \quad
  E^2_{\Lambda{\rm CDM}}(a)
  = \Omega_{m0}\,a^{-3} + (1-\Omega_{m0}).
  \label{eq:E_LCDM}
\end{equation}
%This benchmark expansion history defines the baseline against which all beyond-$\Lambda$CDM deviations are quantified throughout this analysis.
This history of deviations from the reference $\Lambda$CDM model serves as a basis for all beyond $\Lambda$CDM effects measured in this paper.

\subsubsection*{Dynamical dark energy ($w$CDM / CPL)}
The first extension beyond the cosmological constant is to allow the EoS of the DE to be time-dependent, i.e., $w(a)$. For a minimally coupled, barotropic dark energy fluid with pressure $p_{\rm DE} = w(a)\rho_{\rm DE}$, the covariant conservation of the dark energy stress-energy tensor, denoted by $\nabla_\mu T^{\mu\nu}_{\rm DE} = 0$, reduces to the continuity equation \citep{CHEVALLIER_2001,scherrer2015mapping,PhysRevLett.90.091301,cortes2025desi}. 
\begin{equation}
\dot{\rho}_{\rm DE} + 3H\bigl[1 + w(a)\bigr]\rho_{\rm DE} = 0,
\end{equation}
whose solution gives the dark energy density as a function of the scale factor,
\begin{equation}
  \rho_{\rm DE}(a)
  = \rho_{{\rm DE},0}
    \exp\left[
      -3\int_1^a \frac{{\rm d}a'}{a'}\bigl(1+w(a')\bigr)
    \right].
\end{equation}
In the commonly used Chevallier–Polarski–Linder (CPL) parametrization \citep{CHEVALLIER_2001,PhysRevLett.90.091301},
\begin{equation}
  w(a) = w_0 + w_a(1-a)
\end{equation}
where $w_0 \equiv w(a=1)$ is the present-day value and $w_a \equiv -{\rm d}w/{\rm d}a\big|{a=1}$ captures its time evolution. This linear expansion in $(1-a)$ is chosen for its well behaved high redshift limit ($w \to w_0 + w_a$ as $a \to 0$) and can approximate a wide range of scalar field trajectories for quintessence models \citep{Linder:2005in,Caldwell_2005,cortes2025desi}. After the integral is evaluated analytically, the modification function takes the form:
\begin{equation}
  \mathcal{F}_{w{\rm CDM}}(a)
  = a^{-3(1+w_0+w_a)}\,
    \exp\bigl[-3w_a(1-a)\bigr],
\end{equation}
and hence the squared dimensionless Hubble rate,
\begin{equation}
%\begin{aligned}
E^2_{w{\rm CDM}}(a)
  = \Omega_{m0}\,a^{-3} 
  + (1-\Omega_{m0})\,a^{-3(1+w_0+w_a)} \,
    \exp\!\left[
      -3w_a(1-a)
    \right]
%\end{aligned}
\label{eq:E_wCDM}
\end{equation}
Setting $(w_0,w_a) = (-1,0)$ recovers the $\Lambda$CDM limit of Eq.~\eqref{eq:E_LCDM}. Current constraints from DESI BAO combined with CMB and Type Ia supernovae data prefer $w_0 > -1$ and $w_a < 0$ with $2$–$3\sigma$ significance \citep{cortes2025desi}, providing direct motivation to explore dynamical dark energy in weak lensing forecasts \citep{Takada:2003ef}.

\subsubsection*{Interacting dark energy (IDE)}
In interacting dark energy (IDE) scenarios, the dark sector is coupled through an
energy exchange $Q$ between the dark components, while the total stress-energy tensor
remains conserved, $\nabla_\mu\!\left(T^{\mu\nu}_{\rm DM}+T^{\mu\nu}_{\rm DE}\right)=0$.
Following \cite{Benetti:2024dob}, we model the dark sector as a non-adiabatic
generalized Chaplygin gas, in which a vacuum-type component ($w=-1$) exchanges energy
with pressureless cold dark matter, while baryons remain conserved. In a spatially flat
FLRW background the Friedmann and balance equations read
\begin{align}
3H^2 &= \rho_m + \Lambda, \label{eq:IDE_friedmann}\\
\dot{\rho}_{\rm dm} + 3H\rho_{\rm dm} &= \tilde{\Gamma}\,\rho_{\rm dm} = -\dot{\Lambda},
   \label{eq:IDE_continuity}\\
\dot{\rho}_{b} + 3H\rho_{b} &= 0,
\end{align}
where $\rho_m=\rho_{\rm dm}+\rho_b$ is the total matter density and $\tilde{\Gamma}$ is
the dark-matter creation rate (negative for annihilation). Adding the two matter
equations gives the total balance
\begin{equation}
\dot{\rho}_{m} + 3H\rho_{m} = \Gamma\,\rho_{m} = -\dot{\Lambda},
\qquad
\Gamma = \tilde{\Gamma}\,\frac{\rho_{\rm dm}}{\rho_{m}},
\label{eq:IDE_total_balance}
\end{equation}
so that the interaction is conveniently written as
\begin{equation}
Q(a) = \Gamma(a)\,\rho_{m}(a).
\label{eq:Q_Gamma}
\end{equation}

We adopt a power-law running vacuum term,
\begin{equation}
\Lambda(a) = \sigma\,H(a)^{-2\alpha},
\qquad
\sigma = 3\bigl(1-\Omega_{m0}\bigr)\,H_0^{2(\alpha+1)},
\label{eq:Lambda_running}
\end{equation}
where the dimensionless coupling $\alpha>-1$ controls the strength and redshift
dependence of the interaction, and $\sigma$ is fixed by requiring
$\Lambda(a{=}1)=3H_0^2(1-\Omega_{m0})$. This parametrisation contains three natural
cases: the standard cosmological constant ($\alpha=0$), a constant matter-creation
rate ($\alpha=-1/2$), and a creation rate proportional to the expansion rate
($\alpha=-1$). Combining Eqs.~\eqref{eq:IDE_total_balance}--\eqref{eq:Lambda_running}
fixes the effective interaction rate,
\begin{equation}
\Gamma(a) = -\alpha\,\sigma\,H(a)^{-(2\alpha+1)},
\label{eq:Gamma_def}
\end{equation}
so that $\alpha>0$ ($\Gamma<0$) corresponds to an energy flux from dark matter to the
vacuum, and $\alpha<0$ the reverse; both limits vanish as $\alpha\to0$.

Integrating the background equations yields the closed-form expansion history of a
generalised Chaplygin gas,
\begin{equation}
E^2_{\rm IDE}(a) = \left[(1-\Omega_{m0}) + \Omega_{m0}\,a^{-3(1+\alpha)}\right]^{\frac{1}{1+\alpha}},
\label{eq:E_IDE_closed}
\end{equation}
which interpolates between dust-like behaviour at early times and a de~Sitter phase in
the far future. Conserved baryons correspond to the $a^{-3}$ term in the binomial
expansion of Eq.~\eqref{eq:E_IDE_closed}; only cold dark matter exchanges energy with
the vacuum. Casting Eq.~\eqref{eq:E_IDE_closed} in the unified form of
Eq.~\eqref{eq:Ea_unified},
\begin{equation}
E^2_{\rm IDE}(a) = \Omega_{m0}\,a^{-3} + (1-\Omega_{m0})\,\mathcal{F}_{\rm IDE}(a),
\label{eq:E_IDE}
\end{equation}
the modification function is given explicitly by
\begin{equation}
\mathcal{F}_{\rm IDE}(a)
= \frac{1}{1-\Omega_{m0}}
  \left\{\left[(1-\Omega_{m0}) + \Omega_{m0}\,a^{-3(1+\alpha)}\right]^{\frac{1}{1+\alpha}}
  - \Omega_{m0}\,a^{-3}\right\},
\label{eq:F_IDE}
\end{equation}
which satisfies $\mathcal{F}_{\rm IDE}(a)\xrightarrow[\alpha\to0]{}1$, recovering the
$\Lambda$CDM expansion of Eq.~\eqref{eq:E_LCDM} exactly.

The physical consequence of the coupling is that the matter density no longer dilutes
as $a^{-3}$. Using Eqs.~\eqref{eq:IDE_friedmann}, \eqref{eq:Lambda_running} and
\eqref{eq:E_IDE_closed}, one finds
\begin{equation}
\rho_m(a) = 3H_0^2\,\Omega_{m0}\,a^{-3}\,\beta(a),
\qquad
\beta(a) = \frac{a^3}{\Omega_{m0}}\left[E^2_{\rm IDE}(a) - (1-\Omega_{m0})\,E^{-2\alpha}_{\rm IDE}(a)\right],
\label{eq:beta_def}
\end{equation}
with $\beta(a)\to1$ as $\alpha\to0$. The factor $\beta(a)$ quantifies the
creation/annihilation of dark matter and directly controls the suppression
($\alpha<0$) or enhancement ($\alpha>0$) of the matter power spectrum, and hence of the
weak lensing observables, relative to $\Lambda$CDM.

\subsubsection*{Hu--Sawicki $f(R)$ gravity}
In $f(R)$ gravity, the Einstein–Hilbert action is extended by replacing the Ricci scalar $R$ with a general function $R + f(R)$ \citep{Sawicki_2007,DeFelice:2010aj,barausse2008no,nojiri2007introduction},
\begin{equation}
S = \frac{1}{16\pi G}\int {\rm d}^4x\,\sqrt{-g}\,\bigl[R + f(R)\bigr] + S_m.
\label{eq:fR_action}
\end{equation}
Variation with respect to the metric yields fourth-order field equations, which can be recast as second-order equations for the scalar degree of freedom $f_R \equiv {\rm d}f/{\rm d}R$ (the scalaron) via a conformal transformation to the Einstein frame \citep{DeFelice:2010aj,sotiriou2010f}. The Hu-Sawicki model \citep{Hu_2007} adopts the specific functional form 
\begin{equation}
f(R) = -m^2\,\frac{c_1\bigl(R/m^2\bigr)^n}{c_2\bigl(R/m^2\bigr)^n + 1},
\label{eq:HS_fR}
\end{equation}
where $m^2 \equiv H_0^2\,\Omega_{m0}$ sets the mass scale. By requiring that the model reproduces the observed late-time acceleration without a bare cosmological constant, the parameters $(c_1, c_2)$ are related by $c_1/c_2 = 6(1 - \Omega_{m0})/\Omega_{m0}$ \citep{Hu_2007}.
In the high-curvature regime ($R \gg m^2$), the scalaron field value reduces to \citep{Hu:2007nk,Yan:2025nid,Dash:2020yfq,pal2026redshiftspace21cmbispectrummultipoles}
\begin{equation}
f_R(a) \simeq f_{R0}\left[\frac{\Omega_{m0}\,a^{-3} + 4(1 - \Omega_{m0})}{\Omega_{m0} + 4(1 - \Omega_{m0})}\right]^{-(n+1)},
\label{eq:fR_scalaron}
\end{equation}
where $f_{R0} \equiv f_R(a=1)$ is the present-day scalar field amplitude, which serves as the single free parameter of the model (for fixed $n$). Throughout this work we adopt $n = 1$, consistent with the majority of observational analyses \citep{Euclid:2023tqw,Oikonomou:2022wuk,BarrosoVarela:2025zzv}. A key design feature of the Hu–Sawicki model is that the background expansion history is constructed to closely mimic $\Lambda$CDM \citep{Hu:2007nk},
\begin{equation}
E^2_{f(R)}(a) \simeq \Omega_{m0}\,a^{-3} + (1 - \Omega_{m0})
\label{eq:E_fR_LCDMlike}
\end{equation}
so that $\mathcal{F}_{f(R)}(a) \simeq 1$. The primary observational signatures of $f(R)$ gravity therefore arise not from the background geometry but from the scale- and time-dependent enhancement of the effective gravitational coupling on sub-horizon scales, which modifies the growth of perturbations \citep{tsujikawa2008constraints,li2013non,pogosian2008pattern}. In the limit $f_{R0} \to 0$, the scalaron becomes infinitely massive, the Yukawa suppression screens all scales, and GR/$\Lambda$CDM is exactly recovered.

Equations~\eqref{eq:Ea_unified}–\eqref{eq:E_fR_LCDMlike} thus provide a compact, unified description of the background expansion across the beyond-$\Lambda$CDM scenarios that we confront with weak lensing observables in the subsequent sections in the subsequent sections.

\subsection[Linear growth in beyond-$\Lambda$CDM models]%
{Linear growth of structure in beyond-$\Lambda$CDM models}
\label{sec:growth_unified}

In analogy with the background expansion, we adopt a unified description of the linear growth of matter perturbations, highlighting the distinct physical mechanisms at play in each model class. 

The linear growth factor $D(a)$, defined through the matter density contrast $\delta_m(\mathbf{x},a) = D(a)\,\delta_m(\mathbf{x},a_i)$ in the linear regime, encodes how small-amplitude perturbations in the matter density field are amplified by gravitational instability as the Universe expands. Together with the logarithmic growth rate $f(a) \equiv {\rm d}\ln D/{\rm d}\ln a$, it constitutes one of the most discriminating probes of both the expansion history and the underlying theory of gravity \citep{peebles2020large,Linder:2005in,Pogosian:2007sw}. Observational constraints on $f\sigma_8(z) \equiv f(z)\,\sigma_8(z)$ from redshift-space distortions \citep{Feix:2015dla,benisty2021quantifying} and peculiar velocity surveys have demonstrated significant sensitivity to departures from GR, while theoretical analyses have mapped out the growth predictions for quintessence, coupled dark energy, $f(R)$ gravity, and scalar-tensor models \citep{Linder:2005in,Pogosian:2007sw,Gong:2008fh,di2021cosmology}.

In the Newtonian gauge of scalar perturbation theory, the evolution of matter overdensities on sub-horizon scales ($k \gg aH$) is governed by a second-order ordinary differential equation in the scale factor $a$. For all models considered in this work, this equation can be cast in a unified form that explicitly separates the Hubble drag from the gravitational source \citep{Linder:2005in,Ishak:2019aay}:

\begin{equation}
\frac{{\rm d}^2 D(a;k)}{{\rm d}a^2}
+ \mathcal{D}_X(a,k)\,\frac{{\rm d}D(a;k)}{{\rm d}a}
= \mathcal{S}_X(a,k),D(a;k),
\label{eq:growth_unified}
\end{equation}
where $X\in\{\Lambda{\rm CDM}, w{\rm CDM}, {\rm IDE},f(R)\}$ labels the cosmological model.

The coefficient $\mathcal{D}_X(a,k)$ acts as an effective friction term: it encodes the dilution of peculiar velocities by the Hubble flow and is modified whenever the background expansion rate $E(a)$ or its derivative changes relative to $\Lambda$CDM. The coefficient $\mathcal{S}_X(a,k)$ is the gravitational source term: it captures the strength of the effective gravitational coupling driving the collapse of matter perturbations. In GR-based models ($\Lambda$CDM, $w$CDM, IDE), $\mathcal{S}_X$ is scale-independent and set by the Poisson equation, whereas in $f(R)$ gravity it acquires an explicit $k$-dependence through the modified Poisson equation.

\subsubsection*{GR / $\Lambda$CDM}
In GR with a smooth dark energy component (in both $\Lambda$CDM and $w$CDM), the linearized continuity and Euler equations for pressureless matter, combined with the standard Poisson equation $\nabla^2\Phi = 4\pi Ga^2\bar{\rho}_m\delta_m$, yield the well-known growth equation \citep{springel2006large,RevModPhys.75.559,Dodelson:2003ft,Linder:2005in}

\begin{equation}
\frac{{\rm d}^2 D}{{\rm d}a^2}
+ \frac{1}{a}\left[3 + \frac{{\rm d}\ln E(a)}{{\rm d}\ln a}\right]\frac{{\rm d}D}{{\rm d}a}
= \frac{3\Omega_{m0}}{2a^5E^2(a)}\,D.
\label{eq:growth_GR}
\end{equation}
Reading off the drag and source coefficients:
\begin{equation}
\mathcal{D}_{\Lambda\mathrm{CDM}}(a) = \frac{1}{a}\left[3 + \frac{{\rm d}\ln E}{{\rm d}\ln a}\right],
\quad
\mathcal{S}_{\Lambda\mathrm{CDM}}(a) = \frac{3\Omega_{m0}}{2a^5E^2(a)}.
\label{eq:DS_LCDM}
\end{equation}
The drag term depends on the expansion rate only through $E(a)$ and its logarithmic derivative, while the source term is proportional to the matter density fraction $\Omega_m(a) = \Omega_{m0}a^{-3}/E^2(a)$. For $\Lambda$CDM, $E(a)$ is given by Eq.\eqref{eq:E_LCDM}; for $w$CDM/CPL, it is given by Eq.\eqref{eq:E_wCDM}. In both cases the growth factor $D(a)$ is scale-independent, and the growth rate is well approximated by $f(a) \approx \Omega_m(a)^\gamma$ with the growth index $\gamma \simeq 0.55$ for $\Lambda$CDM and $\gamma \simeq 0.55 + 0.05\,[1+w(z=1)]$ for slowly varying $w(a)$ \citep{Linder:2005in,Huterer_2015,WANG20135,Gong_2008}. %CITE WANG 1998

\subsubsection*{Interacting dark energy (IDE)}
In IDE models the energy exchange between the dark sectors modifies the matter
continuity equation and, consequently, the growth of perturbations. Because the
vacuum-type component does not cluster on sub-horizon scales, the gravitational
source is set entirely by the matter density contrast, but the matter density itself
no longer dilutes as $a^{-3}$: from Eqs.~\eqref{eq:Lambda_running} and
\eqref{eq:E_IDE_closed} one has
\begin{equation}
\rho_m(a) = 3H_0^2\,\Omega_{m0}\,a^{-3}\,\beta(a),
\qquad
\beta(a) = \frac{a^3}{\Omega_{m0}}\Big[E^2(a) - (1-\Omega_{m0})\,E^{-2\alpha}(a)\Big],
\label{eq:beta_def}
\end{equation}
where $\beta(a)$ measures the departure of the true matter density from the standard
$a^{-3}$ scaling and satisfies $\beta\to1$ as $\alpha\to0$.

For the interaction $Q=\Gamma(a)\,\rho_m$, the linear growth of the total-matter
contrast $D(a)$ obeys \citep{amendola2004linear,Benetti:2024dob,2009JCAP...07..027C}
\begin{equation}
\begin{split}
   \frac{{\rm d}^2 D}{{\rm d}a^2}
  &+ \left[
       \frac{3}{a}
     + \frac{1}{E(a)}\frac{{\rm d}E(a)}{{\rm d}a}
     + \frac{\Gamma(a)}{a H_0 E(a)}
    \right]\frac{{\rm d}D}{{\rm d}a} \\
  &+ \left[
      \frac{1}{a^2 H_0 E(a)}\frac{{\rm d}\!\left(a\,\Gamma(a)\right)}{{\rm d}a}
    + \frac{\Gamma(a)}{a^2 H_0 E(a)}
    - \frac{3}{2}\,\frac{\Omega_{m0}\,\beta(a)}{a^5 E(a)^2}
    \right]D = 0 ,
\end{split}
\label{eq:growth_IDE}
\end{equation}
where the drag and source coefficients are modified relative to GR,
\begin{align}
  \mathcal{D}_{\rm IDE}(a)
  &= \frac{3}{a}
   + \frac{1}{E(a)}\frac{{\rm d}E(a)}{{\rm d}a}
   + \frac{\Gamma(a)}{a H_0 E(a)},
  \label{eq:D_IDE}\\[4pt]
  \mathcal{S}_{\rm IDE}(a)
  &= \frac{3}{2}\,\frac{\Omega_{m0}\,\beta(a)}{a^5 E(a)^2}
   - \frac{1}{a^2 H_0 E(a)}\frac{{\rm d}\!\left(a\,\Gamma(a)\right)}{{\rm d}a}
   - \frac{\Gamma(a)}{a^2 H_0 E(a)}.
  \label{eq:S_IDE}
\end{align}
The gravitational source in Eq.~\eqref{eq:S_IDE} is proportional to the physical matter
density $\rho_m\propto a^{-3}\beta(a)$, so the creation or annihilation of dark matter
enters growth through the factor $\beta(a)$, while the terms in $\Gamma(a)$ encode the
extra friction and the time dependence of the transfer rate. The dimensionless ratio
$Q/(H\rho_m)=\Gamma/H$ controls the size of the departure from GR growth. For an energy
flux from the vacuum to dark matter ($\Gamma>0$, i.e. $\alpha<0$), $\beta(a)>1$ enhances
the source and the growth rate is boosted relative to $\Lambda$CDM; the opposite occurs
for $\alpha>0$. Since the interaction modifies only the background densities and leaves
the Poisson equation unchanged, the growth factor remains scale-independent,
$D(a;k)\equiv D(a)$. In the non-interacting limit $\alpha\to0$ ($\Gamma\to0$,
$Q\to0$) one has $\beta\to1$, the extra terms vanish, and
$\mathcal{D}_{\rm IDE}\to\mathcal{D}_{\Lambda{\rm CDM}}$,
$\mathcal{S}_{\rm IDE}\to\mathcal{S}_{\Lambda{\rm CDM}}$, recovering
Eq.~\eqref{eq:DS_LCDM}.

\subsubsection*{Hu--Sawicki $f(R)$}
The growth of perturbations in $f(R)$ gravity differs qualitatively from GR-based models because the scalar degree of freedom (the scalaron $f_R$) mediates a fifth force that enhances the effective gravitational coupling on scales below the scalaron Compton wavelength $\lambda_C \sim 1/m(a)$, where $m(a)$ is the scalaron mass $m(a) \equiv (3\,{\rm d}^2f/{\rm d}R^2)^{-1}$ \citep{Hu:2007nk,DeFelice:2010aj,Pogosian:2007sw}.In the quasi-static, sub-horizon limit, the modified Poisson equation can be written as \citep{tsujikawa2008constraints,li2013non,Bean:2006up,Zhao_2014}
\begin{equation}
 -\frac{k^2}{a^2}\,\Phi(k,a) = 4\pi G\,\mu(a,k)\,\bar\rho_m(a)\,\delta_m(k,a),
\label{eq:Poisson_fR}
\end{equation}
where $\Phi$ is the Newtonian potential and $\mu(a,k)$ encodes the scale-dependent modification to the gravitational coupling. In GR/$\Lambda$CDM one has $\mu = 1$, while in Hu–Sawicki $f(R)$ the function $\mu$ is given by \citep{Pogosian:2007sw,Bean:2006up,tsujikawa2008constraints,Zhao_2014}
\begin{equation}
\mu(a,k) = 1 + \frac{1}{3}\,\frac{k^2}{k^2 + a^2\,m^2(a)}.
\label{eq:mu_fR}
\end{equation}
On small scales ($k \gg a,m$), the scalaron propagates freely and $\mu \to 4/3$, enhancing gravity by one-third relative to GR. On large scales ($k \ll a,m$), the scalaron-mediated force is Yukawa-suppressed and $\mu \to 1$, recovering GR.

Incorporating $\mu(a,k)$ into the growth equation yields
\begin{equation}
\frac{{\rm d}^2 D}{{\rm d}a^2}
+ \left[\frac{3}{a} + \frac{1}{E(a)}\frac{{\rm d}E(a)}{{\rm d}a}\right]\frac{{\rm d}D}{{\rm d}a}
= \frac{3}{2}\frac{\Omega_{m0}}{a^5E^2(a)}\mu(a,k)D
\label{eq:growth_fR}
\end{equation}
Since the background expansion closely mimics $\Lambda$CDM (Section~\ref{subsec:background_expansion}), the drag coefficient is unchanged to leading order,
\begin{equation}
\mathcal{D}_{f(R)}(a) \simeq \frac{3}{a} + \frac{1}{E_{\Lambda\rm{CDM}}(a)}\frac{{\rm d}E_{\Lambda\rm{CDM}}(a)}{{\rm d}a},
\label{eq:D_fR}
\end{equation}
and source term
\begin{equation}
\mathcal{S}_{f(R)}(a,k) \simeq \frac{3}{2}\frac{\Omega_{m0}}{a^5E_{\Lambda{\rm CDM}}^2(a)}\mu(a,k).
\label{eq:DS_fR}
\end{equation}
For the Hu–Sawicki model with $n = 1$, the scalaron mass is determined by the background through \citep{Hu:2007nk}
\begin{equation}
m^2(a) = \frac{3H_0^2}{(n+1)\,|f_{R0}|}\,
\frac{\bigl[\Omega_{m0}\,a^{-3} + 4(1-\Omega_{m0})\bigr]^{n+2}}
     {\bigl[\Omega_{m0} + 4(1-\Omega_{m0})\bigr]^{n+1}}.
\label{eq:m_fR}
\end{equation}
Smaller values of $|f_{R0}|$ yield a heavier scalaron, pushing the Compton wavelength to smaller spatial scales and confining the fifth-force enhancement to higher $k$. In the GR limit $f_{R0} \to 0$, the scalaron mass diverges ($m \to \infty$), Eq.\eqref{eq:mu_fR} gives $\mu \to 1$ at all $k$, and the standard $\Lambda$CDM growth of Eq.\eqref{eq:growth_GR} is recovered.

The scale dependence of $\mathcal{S}_{f(R)}$ is the key distinguishing feature of $f(R)$ gravity: it produces a $k$-dependent growth factor $D(a;k)$ that boosts the matter power spectrum on small scales relative to $\Lambda$CDM predictions, while leaving the large-scale growth unchanged. This characteristic signature provides a direct observational handle through weak lensing, which probes the integrated matter distribution across a broad range of scales \citep{Liu_2016, li2013non,Li_2012,Pogosian:2007sw}.

The unified growth equations derived above, Eqs.~\eqref{eq:growth_GR}, \eqref{eq:growth_IDE}, and \eqref{eq:growth_fR}, are each second-order ODEs in the scale factor $a$, and require two initial conditions to specify the growing-mode solution uniquely. We initialise all models at $a_i = 10^{-3}$, deep in the matter-dominated epoch where the dark energy density is negligible ($\Omega_{\rm DE}(a_i) \ll \Omega_m(a_i)$) and the scalaron mass in $f(R)$ is large enough that $\mu(a_i,k) \simeq 1$ for all relevant $k$. In this regime, every model reduces effectively to Einstein–de Sitter (EdS), for which the growing-mode solution is $D(a) \propto a$. We therefore adopt the common initial conditions $D(a_i) = a_i \, \frac{{\rm d}D}{{\rm d}a}\big|_{a=a_i} = 1$ for all scenarios. This choice ensures that differences in the growth factor at later times arise entirely from the distinct drag and source terms of each model, rather than from the initial normalisation.

For GR/$\Lambda$CDM and $w$CDM, the growth factor is scale-independent, $D(a;k) \equiv D(a)$, since both $\mathcal{D}_X$ and $\mathcal{S}_X$ depend only on the background expansion rate $E(a)$. In IDE models, $D(a)$ likewise remains scale-independent: the interaction modifies the matter continuity equation but preserves the standard Poisson equation, so the gravitational source term carries no $k$-dependence. By contrast, in Hu–Sawicki $f(R)$ gravity the factor $\mu(a,k)$ in the source term $\mathcal{S}_{f(R)}$ introduces an explicit scale dependence, and the growth factor must be written as $D(a;k)$ to reflect this.

A useful dimensionless diagnostic of growth is the logarithmic growth rate,
\begin{equation}
f(a;k) \equiv \frac{{\rm d}\ln D(a;k)}{{\rm d}\ln a} = \frac{a}{D(a;k)},\frac{{\rm d}D(a;k)}{{\rm d}a}.
\label{eq:growth_rate}
\end{equation}
In the EdS limit $f = 1$ exactly, while in $\Lambda$CDM the suppression of growth by dark energy gives $f < 1$ at late times. The growth rate is commonly combined with the amplitude of matter fluctuations to form the observable quantity $f\sigma_8(a)$, which can be directly compared with redshift-space distortion (RSD) measurements \citep{Feix:2015dla,benisty2021quantifying}. We define
\begin{equation}
f\sigma_8(a) \equiv f(a)\sigma_8(a)
= \frac{a}{\delta_m(a=1)}\frac{{\rm d}\delta_m(a)}{{\rm d}a}\sigma_{8,0},
\label{eq:fsigma8}
\end{equation}
where $\sigma_{8,0} \equiv \sigma_8(a=1)$ is the present-day normalisation of the linear matter power spectrum smoothed on $8\,h^{-1}{\rm Mpc}$ scales. This quantity is particularly valuable because it is independent of the galaxy bias $b$ in linear theory, making it a clean probe of gravitational physics \citep{Linder:2005in,kazantzidis2018evolution}.

%\begin{figure}[H] % or [tbp]
\begin{center}
\includegraphics[width=0.99\linewidth]{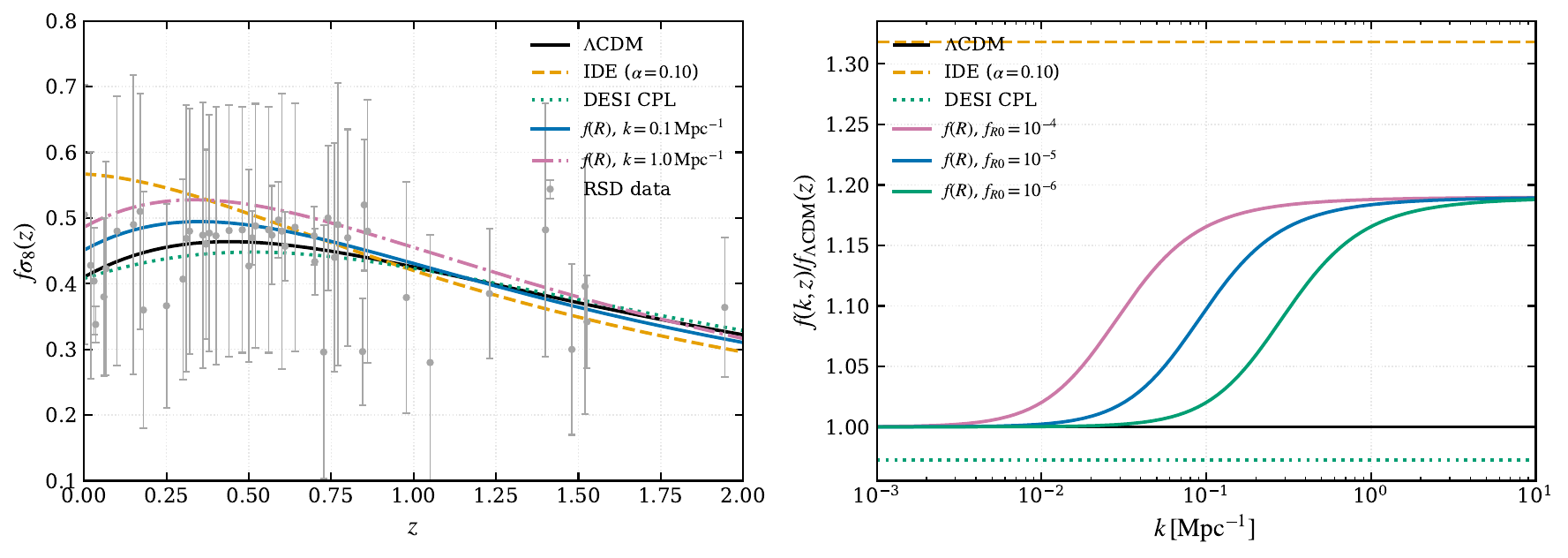}
\captionof{figure}{\textit{Left panel}: Redshift evolution of the linear growth rate $f\sigma_8(z)$ for $\Lambda$CDM (black), the DESI-preferred CPL
parametrisation (green), IDE with coupling $\alpha = 0.10$ (yellow), and Hu--Sawicki $f(R)$ gravity at two representative scales $k = 0.1$ and $1.0\;\mathrm{Mpc}^{-1}$
(blue and pink respectively). Observational data points from redshift-space distortion measurements are taken from \citep{benisty2021quantifying}. 
\textit{Right panel}: Ratio of the growth rate
$f\sigma_8(k,z)$ to its $\Lambda$CDM value at $z = 0.1$ as a function
of $k$, shown for the same set of models and for three
values of the present-day scalaron amplitude,
$|f_{R0}| = 10^{-4},\,10^{-5},\,10^{-6}$.}
\label{fig:growth}
%\end{figure}
\end{center}

Figure~\ref{fig:growth} illustrates the distinct growth signatures predicted by the three beyond-$\Lambda$CDM models described above. The left panel shows the redshift evolution of $f\sigma_8(z)$, with all models evolved from the common EdS initial conditions at $a_i = 10^{-3}$. The curves are indistinguishable during the matter-dominated era and begin to diverge once the dark energy or modified gravity sector becomes dynamically significant at $z \lesssim 2$. In the IDE model, the energy transfer from dark energy to dark matter modifies the effective matter dilution rate, leading to a suppression or enhancement of $f\sigma_8$ depending on the sign and magnitude of the coupling parameter $\alpha$. In Hu–Sawicki $f(R)$, the background expansion tracks $\Lambda$CDM by construction, but the scale-dependent enhancement $\mu(a,k) > 1$ on sub-Compton scales produces a net excess in $f\sigma_8(z)$ even in the linear regime.

The right panel of Figure~\ref{fig:growth} displays the scale dependence of the growth rate at a fixed redshift $z = 0.1$, normalised to the $\Lambda$CDM prediction. For GR-based models ($\Lambda$CDM, $w$CDM, IDE), the ratio is flat across all $k$, confirming the scale-independence of the growth factor in these theories. For Hu–Sawicki $f(R)$, the ratio exhibits a characteristic transition: on scales larger than the Compton wavelength ($k \lesssim a\,m(a)$), $\mu \simeq 1$ and the growth matches $\Lambda$CDM, while on smaller scales ($k \gtrsim a\,m(a)$) the fifth force drives an enhancement that saturates at the $\mu = 4/3$ limit. The transition scale shifts to larger $k$ (smaller spatial scales) as $|f_{R0}|$ decreases, reflecting the heavier scalaron mass and the more efficient Yukawa screening \citep{Tsujikawa:2009ku,Paul:2024lxh}. This distinctive scale-dependent signature is precisely what weak lensing tomography is designed to detect, as the lensing power spectrum integrates over a broad range of $k$ and is therefore sensitive to the transition between the screened and unscreened regimes \citep{Hellwing:2013rxa,Lombriser:2012nn}.

\section[Weak lensing observables and systematics]%
{Weak lensing observables and observational systematics}
\label{sec:wl_observables_systematics}

Having established the background expansion and the linear growth of structure, we now
construct the weak-lensing observables that are our primary probe of the beyond-$\Lambda$CDM
models. Weak lensing traces the intervening matter distribution directly, without
assumptions relating luminous and dark matter \citep{bartelmann2016weak,Kilbinger2015}.
We first introduce the tomographic two-point spectrum and its nonlinear ingredients, then
the two dominant systematics i.e., intrinsic alignments (IA) and photometric-redshift
uncertainties. Throughout we adopt the fiducial cosmology \cite{Planck2018}
\begin{equation*}
\{H_0,\Omega_{m0},\Omega_{b0},A_s,n_s\}=
\{67.7~{\rm km\,s^{-1}Mpc^{-1}},\,0.31,\,0.049,\,2.1\times10^{-9},\,0.965\},
\end{equation*}
and a Vera C.\ Rubin/LSST Year-10 configuration ($f_{\rm sky}=0.4$,
$\bar n_g=26~{\rm arcmin^{-2}}$, $\sigma_\epsilon=0.26$) over $20\le\ell\le1000$ in bins
of $\Delta\ell=50$.

\subsection{Nonlinear matter power spectrum}
\label{sec:pk_nonlinear}
The two-point statistics of the density contrast
$\delta(\mathbf{x},z)\equiv[\rho_m-\bar\rho_m]/\bar\rho_m$ are encoded in the matter power
spectrum,
\begin{equation}
\langle\delta(\mathbf{k},z)\delta^*(\mathbf{k}',z)\rangle=(2\pi)^3\delta_D(\mathbf{k}-\mathbf{k}')\,
P_\delta(k,z).
\label{eq:pk_def}
\end{equation}
In the linear regime the time evolution factorises,
$P^{\rm lin}_\delta(k,z)=[D(a;k)/D(a_{\rm ini};k)]^2P^{\rm lin}_\delta(k,z_{\rm ini})$,
with $D(a;k)$ the growth factor of Sec.~\ref{sec:growth_unified}; it is scale-independent for
$\Lambda$CDM, $w$CDM and IDE, and scale-dependent for $f(R)$.

All spectra are computed with \textsc{camb} \citep{Lewis:2026mif} using the \textsc{HMcode}
nonlinear prescription \citep{Mead:2016zqy}; dynamical dark energy (CPL) uses the PPF
implementation with $w(a)=w_0+w_a(1-a)$. For the GR-based models the scale-independent
growth is read off at a pivot $k_0=0.05~{\rm Mpc^{-1}}$,
$D(z)=[P^{\rm lin}(k_0,z)/P^{\rm lin}(k_0,0)]^{1/2}$. Lacking a calibrated nonlinear
prescription for interacting dark energy, we rescale the $\Lambda$CDM spectrum by the IDE
growth,
\begin{equation}
P^{\rm NL}_{\rm IDE}(k,z)=P^{\rm NL}_{\Lambda\rm CDM}(k,z)
\left[\frac{D_{\rm IDE}(z)}{D_{\Lambda\rm CDM}(z)}\right]^2 ,
\label{eq:pk_ide}
\end{equation}
a leading-order approximation valid where the coupling changes the growth amplitude but
not the mode-coupling shape. Hence, the IDE forecast is therefore exploratory and require $N$-body
calibration \citep{silva2025onelooppowerspectrumcorrections}. For Hu--Sawicki $f(R)$ we use the Gaussian-process
emulator \textsc{FRemu} \citep{Bai:2024cgt}, accurate at percent level over
$0.0089<k/(h\,{\rm Mpc^{-1}})<0.5$, $0<z<3$. Because the Limber projection samples
$k=(\ell+1/2)/\chi$, $\ell_{\max}=1000$ can exceed this range in the lowest-redshift bin;
we therefore use \textsc{FRemu} only within its validity window and fall back outside it to
the growth response
$P^{\rm NL}_{f(R)}=P^{\rm NL}_{\Lambda\rm CDM}[D_{f(R)}(a,k)/D_{\Lambda\rm CDM}(a)]^2$,
which $\to\Lambda$CDM as $|f_{R0}|\to0$. Since \textsc{FRemu} emulates only the power
spectrum, the $f(R)$ bispectrum (Sec.~\ref{sec:bispectrum}) is built from this same
$P^{\rm NL}_{f(R)}$.

\subsection{Tomographic weak-lensing power spectrum}
\label{sec:wl_ps}
The lensing convergence is the line-of-sight projection of the density contrast weighted by
the lensing efficiency of each tomographic bin. Writing the efficiency directly in comoving
distance $\chi$,
\begin{equation}
W_i(\chi)=\frac{3}{2}\,\Omega_{m0}\left(\frac{H_0}{c}\right)^2
\frac{\chi}{a(\chi)}\int_\chi^{\chi_h}\!{\rm d}\chi'\,
n_i(\chi')\,\frac{\chi'-\chi}{\chi'},
\label{eq:lensing_kernel}
\end{equation}
where $n_i(\chi)\,{\rm d}\chi=n_i(z)\,{\rm d}z$ and $\int n_i(z)\,{\rm d}z=1$
(Sec.~\ref{sec:photoz}). Written this way the kernel is dimensionally consistent and
contains no explicit factor of $c/H(z)$ inside the source integral. Under the Limber
and flat-sky approximations ($\ell\gtrsim20$), the tomographic angular power spectrum is
\begin{equation}
C^{ij}_\ell=\int_0^{\chi_h}\!{\rm d}\chi\,
\frac{W_i(\chi)\,W_j(\chi)}{\chi^2}\,
P^{\rm NL}_\delta\!\left(k=\frac{\ell+1/2}{\chi},\,z(\chi)\right),
\label{eq:cl_limber}
\end{equation}
which is manifestly consistent with Eq.~\eqref{eq:lensing_kernel}. In the absence of
baryonic feedback the E-mode shear spectrum equals the convergence spectrum, and we use
$C^{ij}_\ell$ for both. %The full tomographic data vector is shown in Fig.~\ref{fig:datavector}.

\subsection{Intrinsic alignments}
\label{sec:ia}
The observed ellipticity correlation receives four contributions,
\begin{equation}
C^{\epsilon\epsilon}_{ij}(\ell)=C^{GG}_{ij}(\ell)+C^{GI}_{ij}(\ell)+C^{IG}_{ij}(\ell)+C^{II}_{ij}(\ell),
\label{eq:cl_ellipticity}
\end{equation}
where $GG$ is the cosmic-shear signal, the $GI$ cross-terms can partially cancel it, and
$II$ is the intrinsic--intrinsic term. We adopt the nonlinear linear alignment (NLA) model
\cite{Bridle_2007,Hirata:2004gc}, in which
\begin{align}
P_{\delta I}(k,z) &= -\,A_{\rm IA}\,C_1\rho_{\rm cr}\,\frac{\Omega_{m0}}{D(z;k)}\,
P^{\rm NL}_\delta(k,z), \label{eq:pdi}\\
P_{II}(k,z) &= \left[A_{\rm IA}\,C_1\rho_{\rm cr}\,\frac{\Omega_{m0}}{D(z;k)}\right]^2
P^{\rm NL}_\delta(k,z). \label{eq:pii}
\end{align}
Here $A_{\rm IA}$ is a dimensionless amplitude (fiducial $A_{\rm IA}=1$) and
$C_1\rho_{\rm cr}=0.0134$ is the standard NLA normalisation, which already absorbs the
critical density and the physical constant
$C_1=5\times10^{-14}\,h^{-2}M_\odot^{-1}{\rm Mpc^3}$. The alignment contributions enter
Eq.~\eqref{eq:cl_limber} through the intrinsic kernel
$W_i^{\rm IA}(\chi)=n_i(\chi)\,f_{\rm IA}(z;k)$ with
$f_{\rm IA}(z;k)=-A_{\rm IA}C_1\rho_{\rm cr}\Omega_{m0}/D(z;k)$. For $\Lambda$CDM, $w$CDM and
IDE, $f_{\rm IA}$ depends only on $z$; for $f(R)$ the scale-dependent $D(z;k)$ makes the
alignment amplitude $k$-dependent, and the kernel is rebuilt at each multipole with
$k=(\ell+1/2)/\chi$.

\subsection{Photometric redshifts and tomographic binning}
\label{sec:photoz}
\begin{figure}[t]
\centering
\includegraphics[width=0.49\textwidth]{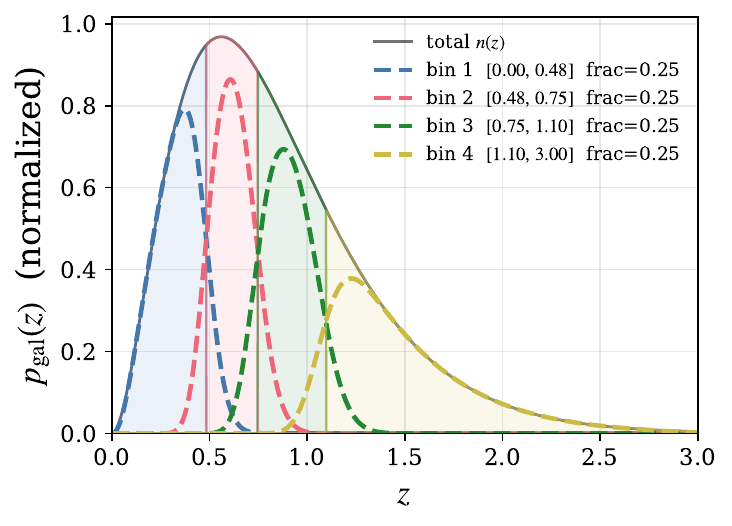}\hfill
\includegraphics[width=0.49\textwidth]{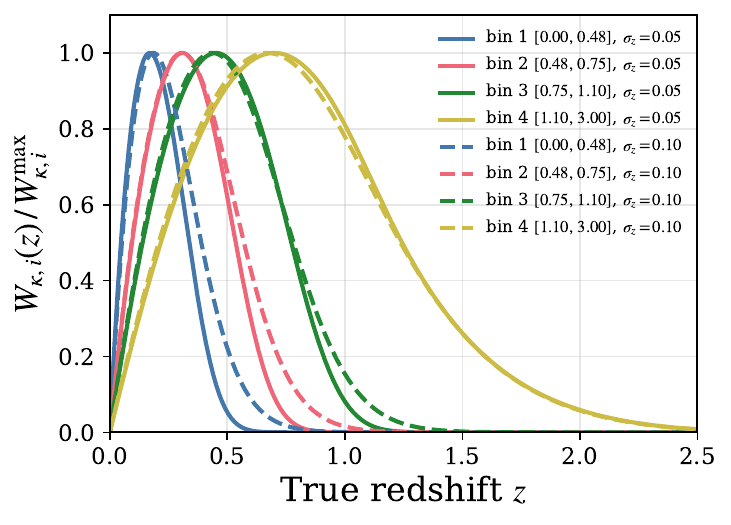}
\caption{\emph{Left:} LSST gold $n(z)$ (Eq.~\ref{eq:nz_gold}) split into four
equal-number photometric bins via the Bayes-consistent convolution of
Eq.~\eqref{eq:nz_convolved}. \emph{Right:} the corresponding peak-normalised lensing
kernels $W_i(z)$ (Eq.~\ref{eq:lensing_kernel}); dashed curves show the pessimistic
$\sigma_z=0.10$ case, which broadens the bins and smooths the kernels, degrading radial
resolution.}
\label{fig:dist-kernel}
\end{figure}

We use the LSST ``gold'' sample, with true-redshift distribution
\begin{equation}
n(z)=\frac{1}{2z_0}\left(\frac{z}{z_0}\right)^2\exp\!\left[-\frac{z}{z_0}\right],
\qquad z_0=0.28,\quad \int_0^\infty n(z)\,{\rm d}z=1,
\label{eq:nz_gold}
\end{equation}
and Gaussian photometric scatter
\begin{equation}
p(z_{\rm ph}\mid z)=\frac{1}{\sqrt{2\pi}\,\sigma_z(1+z)}
\exp\!\left[-\frac{(z_{\rm ph}-z)^2}{2\sigma_z^2(1+z)^2}\right],
\label{eq:pzph}
\end{equation}
with fiducial $\sigma_z=0.05$ and a pessimistic $\sigma_z=0.10$. The effective distribution
of photometric bin $i=[z^{(i)}_{\rm ph},z^{(i+1)}_{\rm ph}]$ is built the Bayes-consistent
way, weighting $n(z)$ by the probability of scatter into the bin,
\begin{equation}
n_i(z)=\frac{n(z)\displaystyle\int_{z^{(i)}_{\rm ph}}^{z^{(i+1)}_{\rm ph}}\!
{\rm d}z_{\rm ph}\,p(z_{\rm ph}\mid z)}
{\displaystyle\int\!{\rm d}z\,n(z)\int_{z^{(i)}_{\rm ph}}^{z^{(i+1)}_{\rm ph}}\!
{\rm d}z_{\rm ph}\,p(z_{\rm ph}\mid z)},
\label{eq:nz_convolved}
\end{equation}
the inner integral being
$\tfrac12[\mathrm{erf}((z^{(i+1)}_{\rm ph}-z)/\sqrt2\sigma)-\mathrm{erf}((z^{(i)}_{\rm ph}-z)/\sqrt2\sigma)]$,
$\sigma=\sigma_z(1+z)$. We use $N_{\rm tomo}=4$ contiguous bins with edges chosen to hold
equal galaxy numbers, $z_{\rm ph}\in\{0,0.48,0.75,1.10,3.0\}$ for the fiducial $n(z)$. The
per-bin density is $\bar n_i=f_i\bar n_g$ with $f_i$ the galaxy fraction, and the shape-noise
level is $\mathcal{N}_i=\sigma_\epsilon^2/\bar n_i$. Figure~\ref{fig:dist-kernel} shows the
resulting bins and lensing kernels.

\subsection{Covariance and scope}
\label{sec:covariance}
We use the Gaussian power spectrum covariance, diagonal in $\ell$,
\begin{equation}
{\rm Cov}\big[C^{ij}_\ell,C^{mn}_\ell\big]
=\frac{\tilde C^{im}_\ell\tilde C^{jn}_\ell+\tilde C^{in}_\ell\tilde C^{jm}_\ell}
{(2\ell+1)\,f_{\rm sky}\,\Delta\ell},
\qquad \tilde C^{ij}_\ell=C^{ij}_{\ell,\rm obs}+\delta_{ij}\mathcal{N}_i.
\label{eq:cov_ps}
\end{equation}
We deliberately neglect the connected non-Gaussian and super-sample terms; the forecasts are
therefore exploratory at $\ell\lesssim1000$. Consistently, the bispectrum covariance is taken at Gaussian order (Sec.~\ref{sec:bispectrum}), and the power spectrum--bispectrum
cross-covariance, which vanishes at Gaussian order, is neglected, so the combined Fisher
matrix is $F_{\rm PS+BS}=F_{\rm PS}+F_{\rm BS}$.

\section[Weak lensing bispectrum]%
        {Weak lensing bispectrum formalism}
%\label{sec:three_point}
\label{sec:bispectrum}

The two-point power spectrum captures only the Gaussian component of the convergence field. However, gravitational evolution is intrinsically nonlinear: mode coupling during structure formation generates non-Gaussian features in the matter density field and are more prominent at late times and on small scales \citep{bernardeau2002large, cooray2001weak}. The lowest-order statistic sensitive to this non-Gaussianity is the three-point function, or equivalently, the bispectrum in harmonic space. The weak lensing bispectrum therefore carries information about the nonlinear growth of structure that is largely complementary to and statistically independent of the power spectrum \citep{Takada:2003ef, Kayo:2013aha, Takada_2003, Valageas_2012}.

From a cosmological standpoint, the bispectrum is particularly valuable for breaking parameter degeneracies that plague power spectrum-only analyses. For instance, the amplitude of the matter fluctuations $\sigma_8$ and the matter density $\Omega_m$ enter the power spectrum primarily through the combination $S_8 \equiv \sigma_8(\Omega_m/0.3)^{0.5}$, leaving a well-known banana-shaped degeneracy in the $(\Omega_m, \sigma_8)$ plane. The bispectrum, being sensitive to the skewness of the density field, responds to a different combination of these parameters and can significantly tighten constraints when combined with the power spectrum \citep{Takada:2003ef, Kayo:2013aha, Coulton_2019}. In the context of beyond-$\Lambda$CDM physics, where modified gravity or dark-sector interactions alter the nonlinear growth in model-specific ways, the bispectrum provides an additional lever arm for distinguishing competing scenarios \citep{2012JCAP...02..047G, Gil_Mar_n_2016, Yankelevich_2018, 2021JCAP...07..008G,Gil_Mar_n_2011}.

In this section, we describe the modeling of the three-dimensional matter bispectrum and its projection onto the sky as the tomographic weak lensing bispectrum.

\begin{figure}[t]
\centering
\includegraphics[width=0.9\textwidth]{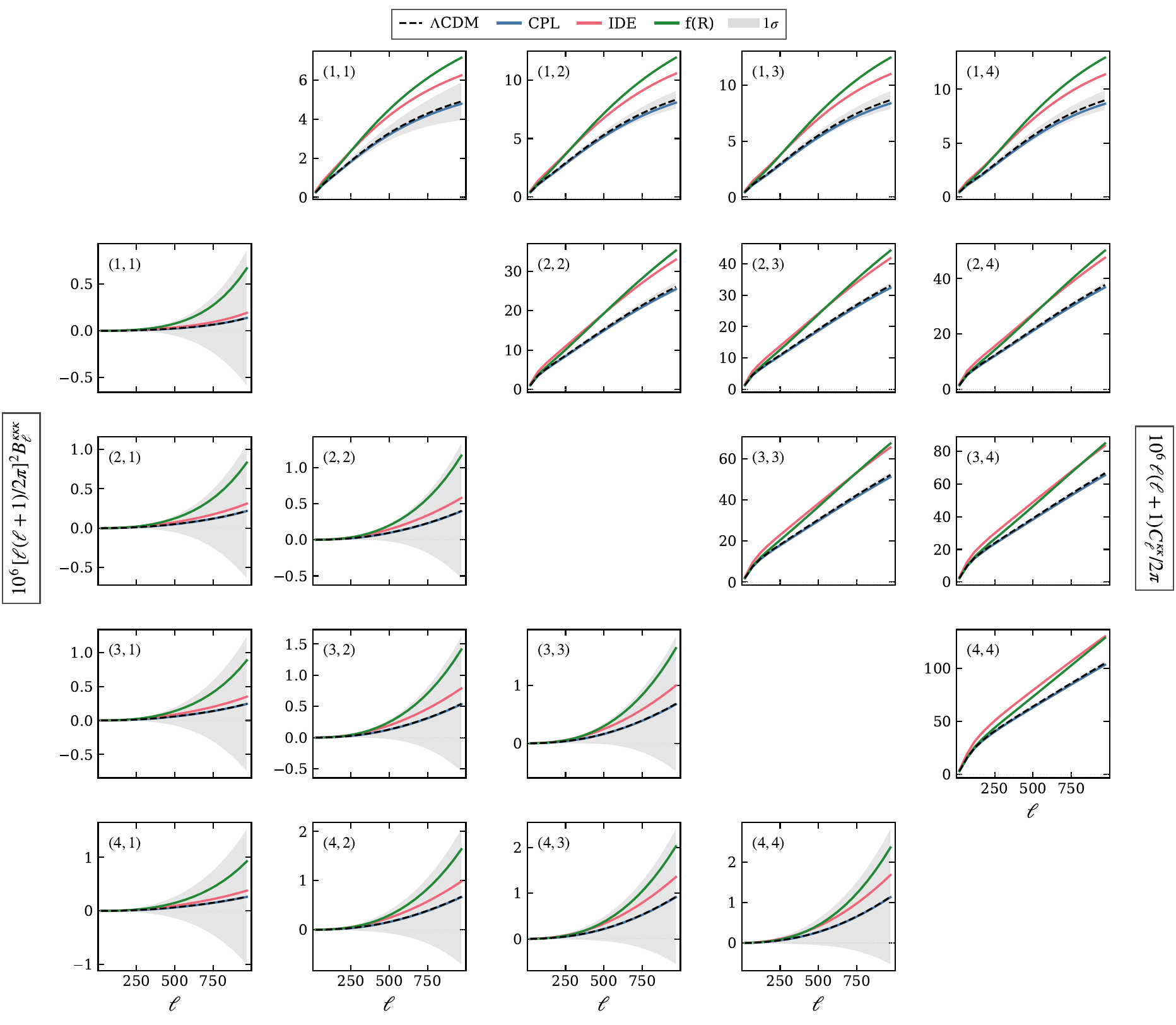}
\caption{Tomographic weak-lensing data vector entering the Fisher analysis for the four
equal-number LSST-Y10 bins. Upper triangle: convergence power spectra (auto and cross); lower triangle: the corresponding
(equilateral) convergence bispectra. Curves compare $\Lambda$CDM, CPL, IDE and Hu--Sawicki
$f(R)$; the model-specific departures are what the joint PS+BS vector is designed to detect.}
\label{fig:datavector}
\end{figure}

\subsection{The matter bispectrum}
\label{sec:matter_bispectrum}
The matter bispectrum $B_\delta$ is the three-point analogue of the power spectrum,
\begin{equation}
\langle\delta(\mathbf{k}_1)\delta(\mathbf{k}_2)\delta(\mathbf{k}_3)\rangle
=(2\pi)^3\delta_D(\mathbf{k}_1+\mathbf{k}_2+\mathbf{k}_3)\,
B_\delta(k_1,k_2,k_3),
\label{eq:bk_def}
\end{equation}
non-zero only for closed triangles $\mathbf{k}_1+\mathbf{k}_2+\mathbf{k}_3=0$. At
tree level, second-order perturbation theory gives
$B_\delta=2F_2(\mathbf{k}_1,\mathbf{k}_2)P^{\rm lin}(k_1)P^{\rm lin}(k_2)+{\rm 2\ perm.}$,
with the standard mode-coupling kernel
\begin{equation}
F_2(\mathbf{k}_1,\mathbf{k}_2)=\frac{5}{7}
+\frac{1}{2}\cos\theta_{12}\!\left(\frac{k_1}{k_2}+\frac{k_2}{k_1}\right)
+\frac{2}{7}\cos^2\theta_{12},
\qquad \cos\theta_{12}=\hat{\mathbf{k}}_1\!\cdot\!\hat{\mathbf{k}}_2 .
\label{eq:f2_tree}
\end{equation}
To capture nonlinear evolution we use the effective kernel of
\cite{2012JCAP...02..047G, Gil_Mar_n_2016}, which replaces the tree-level constants with
scale- and time-dependent fitting functions calibrated on $N$-body simulations,
\begin{equation}
F_2^{\rm eff}(\mathbf{k}_1,\mathbf{k}_2)=\frac{5}{7}\,a(n_1,k_1)a(n_2,k_2)
+\frac{1}{2}\cos\theta_{12}\!\left(\frac{k_1}{k_2}+\frac{k_2}{k_1}\right)b(n_1,k_1)b(n_2,k_2)
+\frac{2}{7}\cos^2\theta_{12}\,c(n_1,k_1)c(n_2,k_2),
\label{eq:f2_eff}
\end{equation}
so that $B_\delta(k_1,k_2,k_3)=2F_2^{\rm eff}(\mathbf{k}_1,\mathbf{k}_2)
P^{\rm NL}(k_1)P^{\rm NL}(k_2)+{\rm 2\ perm.}$, evaluated with the nonlinear power
spectra of Sec.~\ref{sec:pk_nonlinear}. The functions are
\begin{align}
a(n,k)&=\frac{1+\sigma_8^{a_6}(z)\,[0.7\,Q_3(n)]^{1/2}(q\,a_1)^{n+a_2}}{1+(q\,a_1)^{n+a_2}},\\
b(n,k)&=\frac{1+0.2\,a_3(n+3)(q\,a_7)^{n+3+a_8}}{1+(q\,a_7)^{n+3.5+a_8}},\\
c(n,k)&=\frac{1+\dfrac{4.5\,a_4}{1.5+(n+3)^4}(q\,a_5)^{n+3+a_9}}{1+(q\,a_5)^{n+3.5+a_9}},
\label{eq:abc}
\end{align}
with $Q_3(n)=(4-2^n)/(1+2^{n+1})$, the reduced wavenumber $q=k/k_{\rm nl}(z)$ defined by
$k_{\rm nl}^3P^{\rm lin}(k_{\rm nl},z)/(2\pi^2)=1$, and the effective slope
$n\equiv n_{\rm eff}(k,z)={\rm d}\ln P^{\rm lin}/{\rm d}\ln k$. We adopt the calibrated
coefficients \citep{2012JCAP...02..047G}
\begin{equation*}
(a_1,\dots,a_9)=(0.484,\,3.740,\,-0.849,\,0.392,\,1.013,\,-0.575,\,0.128,\,-0.722,\,-0.926).
\end{equation*}
The fitting formula was calibrated for $\Lambda$CDM over $0.03\lesssim k/(h\,{\rm Mpc^{-1}})\lesssim0.4$,
$z\lesssim1.5$. We apply it to the beyond-$\Lambda$CDM models by evaluating
Eqs.~\eqref{eq:f2_eff}--\eqref{eq:abc} with the corresponding $P^{\rm NL}(k,z)$; this is a
leading-order approximation, since the mode-coupling kernel is not recalibrated for the
modified dynamics, and we treat the bispectrum forecasts as exploratory (Sec.~\ref{sec:results}).

\subsection{Projected convergence bispectrum}
\label{sec:kappa_bispectrum}
Under the Limber approximation the tomographic convergence bispectrum is the line-of-sight
projection of $B_\delta$ weighted by three lensing kernels,
\begin{equation}
B^{\kappa}_{ijk}(\ell_1,\ell_2,\ell_3)=\int_0^{\chi_h}\!{\rm d}\chi\,
\frac{W_i(\chi)\,W_j(\chi)\,W_k(\chi)}{\chi^4}\,
B_\delta\!\left(\frac{\ell_1}{\chi},\frac{\ell_2}{\chi},\frac{\ell_3}{\chi};z(\chi)\right),
\label{eq:bkappa_limber}
\end{equation}
with $W_i$ the kernel of Eq.~\eqref{eq:lensing_kernel} and the harmonic-space triangle
condition $\boldsymbol{\ell}_1+\boldsymbol{\ell}_2+\boldsymbol{\ell}_3=0$. Although
Fig.~\ref{fig:datavector} displays equilateral configurations for clarity, the Fisher
analysis uses \emph{all} triangle configurations with $\ell_1\le\ell_2\le\ell_3$ in
$20\le\ell\le1000$ ($\Delta\ell=50$) that satisfy $|\ell_1-\ell_2|\le\ell_3\le\ell_1+\ell_2$,
together with every tomographic triplet $(i,j,k)$. The complete tomographic data vector for power spectrum and bispectrum case is shown in Fig.~\ref{fig:datavector}. The upper triangle shows the tomographic convergence power spectra (auto-spectra on the diagonal, cross-spectra off it) for the four equal-number LSST-Y10 bins. And the lower triangle corresponds to equilateral convergence bispectra for the same bin combinations. In each panel we overplot $\Lambda$CDM, CPL, IDE and
Hu--Sawicki $f(R)$, so that the small model-dependent departures from $\Lambda$CDM are visible in both statistics. These departures and the extra, differently-weighted information carried by the three-point signal are precisely what the joint PS+BS data vector is constructed to detect.

\subsection{Bispectrum covariance and Fisher forecast}
\label{sec:bs_fisher}
We adopt the Gaussian bispectrum covariance of \cite{Kayo:2013aha}, which is
block-diagonal in triangle configuration,
\begin{equation}
{\rm Cov}\big[B^{\kappa}(\ell_1,\ell_2,\ell_3)\big]
=\frac{\Omega_s\,\Delta_{\ell_1\ell_2\ell_3}}{N_{\rm tri}(\ell_1,\ell_2,\ell_3)}\,
\tilde C_{\ell_1}\tilde C_{\ell_2}\tilde C_{\ell_3},
\label{eq:bs_cov}
\end{equation}
where $\tilde C_\ell$ is the observed spectrum including shape noise, $\Delta=1,2,6$ for
scalene, isosceles and equilateral triangles respectively, and the number of triangle
configurations in a bin is
\begin{equation}
N_{\rm tri}(\ell_1,\ell_2,\ell_3)=\frac{\Omega_s^2\,\ell_1\ell_2\ell_3\,(\Delta\ell)^3}
{(2\pi)^3\,\sqrt{2\ell_1^2\ell_2^2+2\ell_2^2\ell_3^2+2\ell_3^2\ell_1^2-\ell_1^4-\ell_2^4-\ell_3^4}},
\label{eq:ntri}
\end{equation}
the radical being $4\times$ the area of the triangle with sides $(\ell_1,\ell_2,\ell_3)$.
Consistent with Sec.~\ref{sec:covariance}, we retain only the Gaussian term and neglect the
power spectrum--bispectrum cross-covariance, which vanishes at this order.

Parameter constraints follow from the Fisher matrix. For a data vector $\mathbf{D}^X$
($X={\rm PS,BS}$) with covariance $\mathsf{C}^X$,
\begin{equation}
F^X_{\alpha\beta}=\sum
\frac{\partial \mathbf{D}^X}{\partial\theta_\alpha}\,
\big(\mathsf{C}^X\big)^{-1}\,
\frac{\partial \mathbf{D}^X}{\partial\theta_\beta},
\label{eq:fisher}
\end{equation}
with derivatives evaluated by central finite differences about the fiducial cosmology. We
vary the full parameter set, the model parameter(s) plus
$\{\Omega_{m0},H_0,n_s,A_s,\sigma_z,A_{\rm IA}\}$ and impose Gaussian
Planck priors on the shared cosmological parameters. Since the Gaussian PS and BS
covariances are uncorrelated, the combined constraint is
$F_{\rm PS+BS}=F^{\rm PS}+F^{\rm BS}+F^{\rm prior}$, and marginalized errors are
$\sigma(\theta_\alpha)=[(F^{-1})_{\alpha\alpha}]^{1/2}$.

\begin{figure}[t]
\centering
\includegraphics[width=0.45\textwidth]{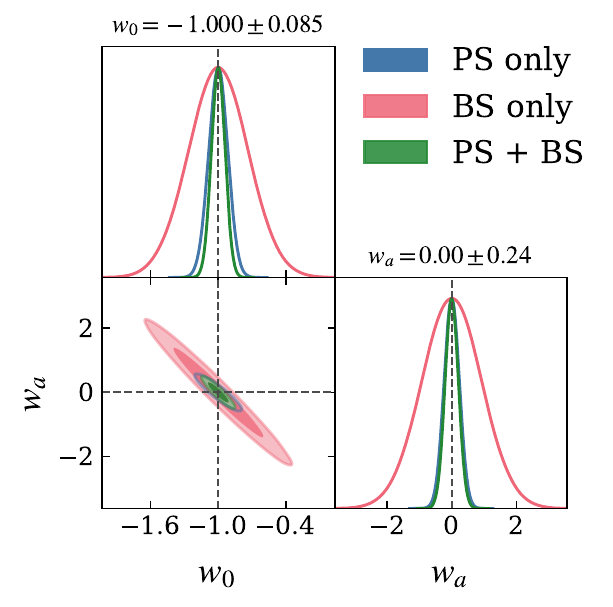}
\includegraphics[width=0.45\textwidth]{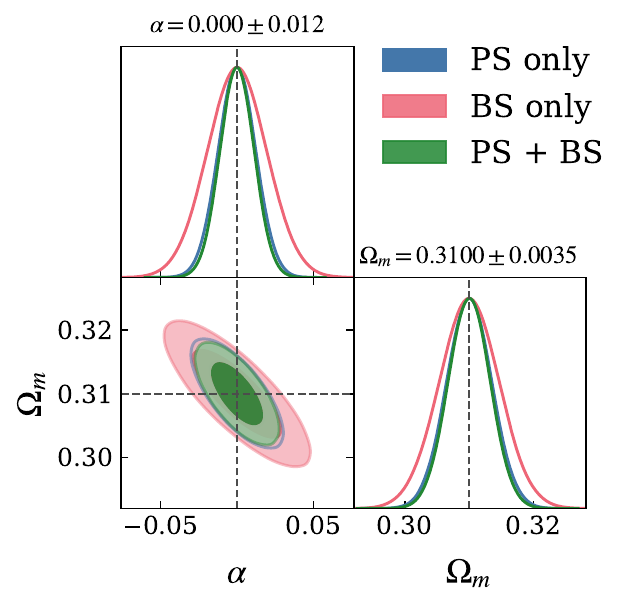}
\includegraphics[width=0.45\textwidth]{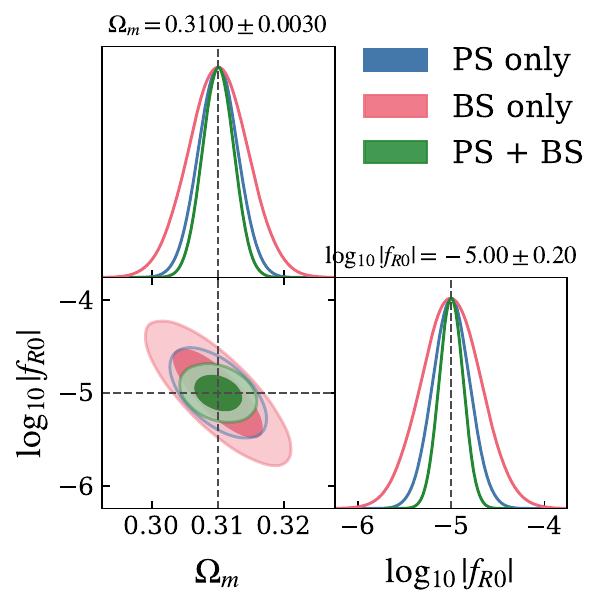}
\caption{Marginalized $68\%$ and $95\%$ confidence contours for the three
beyond-$\Lambda$CDM models. \emph{Left:} CPL $(w_0,w_a)$;
\emph{centre:} IDE $(\alpha,\Omega_{m0})$;
\emph{right:} Hu--Sawicki $(\log_{10}|f_{R0}|,\Omega_{m0})$.
In each panel the wider contour is the power spectrum-only forecast and the
tighter one the joint PS+BS result; the bispectrum rotates and shrinks the
degeneracy ellipse, most strongly for CPL and $f(R)$.}
\label{fig:contours}
\end{figure}

\section{Forecast results}
\label{sec:results}

We now present the LSST-Y10 forecasts obtained from the Fisher matrix of
Eq.~\eqref{eq:fisher}, comparing the power spectrum only case (PS) with the
joint power spectrum plus bispectrum analysis (PS+BS). All constraints
marginalize over the shared set $\{\Omega_{m0},H_0,n_s,A_s,\sigma_z,A_{\rm IA}\}$
with the Planck priors of Sec.~\ref{sec:bs_fisher}, and vary the
model parameter(s) on top of that set. Unless stated otherwise we quote
$1\sigma$ marginalized errors over $20\le\ell\le1000$ ($\Delta\ell=50$) for the
four chosen tomographic bins.

\begin{table}[t]
\centering
\begin{tabular}{llccc}
\hline\hline
Model & Parameter & PS & BS & PS+BS \\
\hline
CPL & $\sigma(w_0)$ & $0.085$ & $0.265$ & $0.061$ \\
    & $\sigma(w_a)$ & $0.237$ & $0.924$ & $0.208$ \\
    & $\sigma(\Omega_{m0})$ & $0.0039$ & $0.0042$ & $0.0032$ \\
    & ${\rm FoM}(w_0,w_a)$ & $119$ & $19.8$ & $207$ \\
\hline
IDE & $\sigma(\alpha)$ & $0.0123$ & $0.0196$ & $0.0112$ \\
    & $\sigma(\Omega_{m0})$ & $0.0035$ & $0.0047$ & $0.0033$ \\
    & ${\rm FoM}(\alpha,\Omega_{m0})$ & $3.32\times10^{4}$ & $1.82\times10^{4}$ & $3.73\times10^{4}$ \\
\hline
$f(R)$ & $\sigma(\log_{10}|f_{R0}|)$ & $0.198$ & $0.320$ & $0.129$ \\
       & $\sigma(\Omega_{m0})$ & $0.0030$ & $0.0045$ & $0.0024$ \\
       & ${\rm FoM}(\log_{10}|f_{R0}|,\Omega_{m0})$ & $2053$ & $1121$ & $3364$ \\
\hline\hline
\end{tabular}
\caption{Marginalized $1\sigma$ forecast constraints under the LSST-Y10
configuration for the power spectrum only (PS), bispectrum-only (BS), and joint
(PS+BS) data vectors, after marginalising over the shared parameters
$\{\Omega_{m0},H_0,n_s,A_s,\sigma_z,A_{\rm IA}\}$. The figure of merit is
${\rm FoM}=[\det\mathsf{C}_{2\times2}]^{-1/2}$ in the plane indicated. Adding
the bispectrum improves the FoM by factors of $1.74$ (CPL), $1.64$ ($f(R)$) and
$1.12$ (IDE).}
\label{tab:forecasts}
\end{table}

\begin{figure}[t]
\centering
\includegraphics[width=0.9\textwidth]{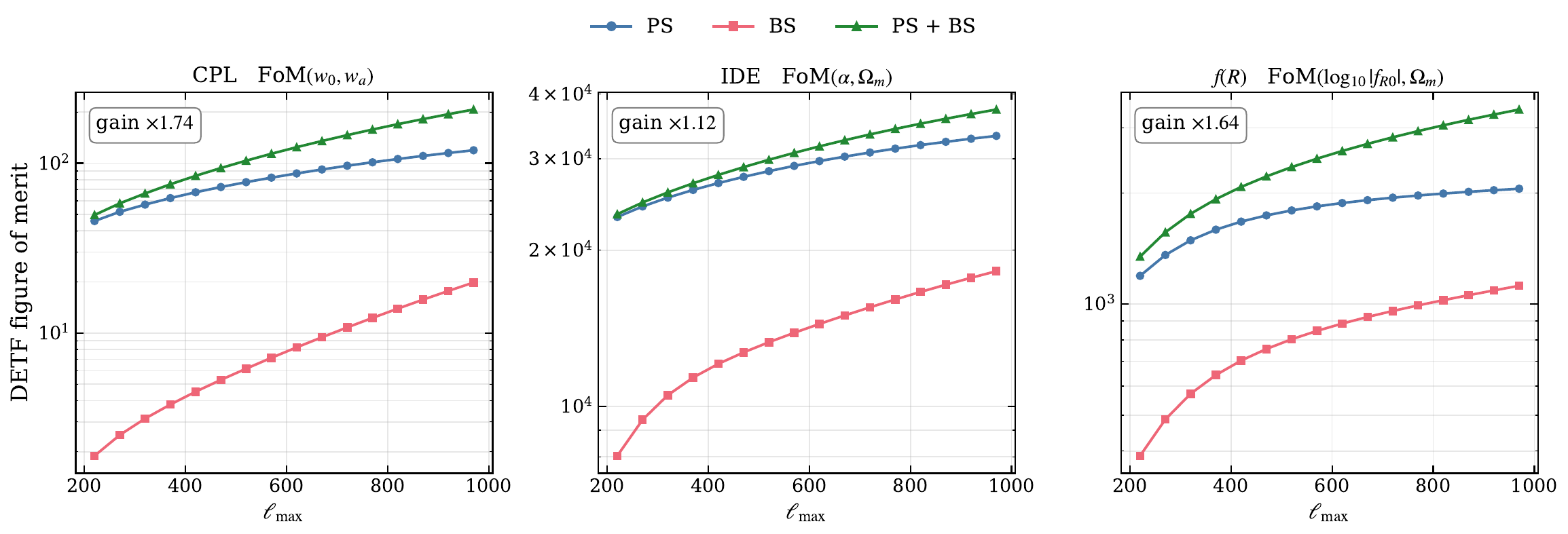}
\caption{Two-parameter figure of merit as a function of the maximum
multipole $\ell_{\max}$ for the three models, for the PS-only (dashed) and
PS+BS (solid) data vectors. The bispectrum contribution grows with
$\ell_{\max}$ as progressively more nonlinear, mode-coupled scales enter the
data vector, saturating near $\ell_{\max}=1000$ where our Gaussian-covariance
and kernel approximations begin to break down.}
\label{fig:fom_lmax}
\end{figure}

Table~\ref{tab:forecasts} collects the marginalized errors and the
figure of merit (FoM) gains, and Figs.~\ref{fig:contours}--\ref{fig:fom_lmax}
display the corresponding confidence contours and their scale dependence.

\paragraph{Dynamical dark energy (CPL):}
The bispectrum has its largest impact on the CPL parametrization. Adding the
three-point information tightens the equation of state constraints from
$\sigma(w_0)=0.085$, $\sigma(w_a)=0.237$ to $\sigma(w_0)=0.061$,
$\sigma(w_a)=0.208$, improving the $(w_0,w_a)$ figure of merit from $119$ to
$207$ (a factor of $1.74$; left panel of Fig.~\ref{fig:contours}). The gain
reflects the different $(w_0,w_a)$ degeneracy direction picked out by the
mode-coupling kernel: the bispectrum responds to the growth history through
$F_2^{\rm eff}$ and $n_{\rm eff}$, partially rotating the elongated
power spectrum ellipse.

\paragraph{Interacting dark energy (IDE):}
For the coupling $\alpha$ the bispectrum adds comparatively little,
$\sigma(\alpha)=0.0123\to0.0112$ and $\sigma(\Omega_{m0})=0.0035\to0.0033$, a
figure of merit gain of $1.12$ (middle panel of Fig.~\ref{fig:contours}). This
is expected: in our leading-order treatment the coupling enters the nonlinear
spectrum only through the growth rescaling of Eq.~\eqref{eq:pk_ide}, so the
same information largely propagates into both statistics. We therefore regard
the IDE bispectrum forecast as exploratory, pending an $N$-body calibration of
the mode-coupling kernel for interacting models.

\paragraph{Hu--Sawicki $f(R)$:}
The scale-dependent growth of $f(R)$ leaves a distinctive imprint on the
bispectrum, and the joint analysis reaches $\sigma(\log_{10}|f_{R0}|)=0.129$ and
$\sigma(\Omega_{m0})=0.0024$, improving the figure of merit from $2053$ to
$3364$ (a factor of $1.64$; right panel of Fig.~\ref{fig:contours}). Because the
fifth force enhances mode coupling on sub-Compton scales precisely where the
bispectrum signal peaks, the three-point statistic breaks the
$\log_{10}|f_{R0}|$--$\Omega_{m0}$ degeneracy more effectively than the power
spectrum alone.

Two caveats to this analysis are worth noting. First, our bispectrum covariance is
Gaussian; \cite{Kayo:2013aha} showed that including the non-Gaussian and
super-sample (halo sample variance) terms degrades the bispectrum more than the
power spectrum, cutting the apparent improvement of a Gaussian-only forecast
from a factor of two or three down to the ten-to-twenty-per-cent level. Our
CPL and IDE gains should thus be read as optimistic upper bounds. Second, the
relative benefit of the bispectrum diminishes as tomographic bins are added,
since the power spectrum then already recovers much of the growth information;
with four bins the residual gain from the three-point statistic is
correspondingly reduced.

A feature common to all three models is that the bispectrum alone
constrains far more weakly than the power spectrum alone (the ``BS only''
contours in Fig.~\ref{fig:contours} are in every case the widest). This is
expected, and worth making explicit because the literature is not unanimous.
\cite{Kayo:2013aha} found the same hierarchy in their tomographic forecast,
with a bispectrum-only figure of merit roughly half the power spectrum-only value, and gave the physical reason: large-scale structure grows from a nearly
Gaussian initial field, so the power spectrum carries the primary cosmological information, while the bispectrum only measures the secondary non-Gaussianity generated later by nonlinear mode coupling. The three-point statistic is therefore intrinsically information-poor on its own, and becomes
valuable chiefly as a complement that breaks degeneracies when added to the
power spectrum which is exactly the behavior we find.

Three features of our set-up deepen this in our case. First, we restrict the
analysis to $\ell\le1000$: the bispectrum signal-to-noise rises steeply toward
smaller, more nonlinear scales, so a conservative $\ell_{\max}$ chosen to stay
within the validity of our nonlinear prescriptions deliberately excludes the
high-$\ell$ regime where the bispectrum-alone would be most powerful
(Fig.~\ref{fig:fom_lmax}). Second, the beyond-$\Lambda$CDM physics enters our
bispectrum only through the growth-rescaled nonlinear power spectrum, with a
$\Lambda$CDM-calibrated mode-coupling kernel; any genuinely enhanced
three-point response of the modified dynamics of the kind that made the
bispectrum competitive in the simulation-based analysis of \cite{Sato:2013mq}, which \cite{Kayo:2013aha} could not reproduce is by construction
absent. Third, with shape noise entering each factor of the bispectrum
covariance, the three-point measurement becomes noise-dominated over
$20\le\ell\le1000$ more rapidly than the two-point one. The bispectrum-alone is
thus not a competitive stand-alone probe in our configuration; its role, and
the result we report, is the degeneracy-breaking gain it provides in
combination with the power spectrum.

To summarize our results are best understood as a controlled lower
bound on the value of the weak-lensing bispectrum rather than its ceiling.
Even within these conservative choices, a Gaussian covariance, a
$\Lambda$CDM-calibrated kernel, and a linear-to-mildly-nonlinear cut at
$\ell_{\max}=1000$, the three-point statistic delivers a real,
model-dependent tightening of the constraints when combined with the power spectrum, and it does so precisely where the new physics reshapes nonlinear
growth. Relaxing each assumption should only increase this return: extending
to smaller scales with a jointly modeled baryonic sector would add the
high-$\ell$ triangles that carry most of the bispectrum signal, a non-Gaussian
and super-sample covariance would place the combination on a realistic error
budget, and a mode-coupling kernel recalibrated on beyond-$\Lambda$CDM $N$-body
simulations particularly for the interacting and $f(R)$ sectors, where our
growth-only treatment is most limiting would capture three-point
information that the present analysis discards by construction. Each of these
is a concrete direction for future work, and together they set the agenda for turning the forecasts presented here into a competitive bispectrum analysis for
Stage~IV data.

\section{Conclusions}
\label{sec:conclusions}

We have forecast the constraining power of the tomographic weak-lensing power
spectrum and bispectrum for three representative departures from $\Lambda$CDM, CPL dynamical dark energy, interacting dark energy, and Hu--Sawicki
$f(R)$ gravity under a Vera~C.~Rubin/LSST Year-10 configuration. For each
model we built the background expansion and linear growth consistently
(Sec.~\ref{sec:de_models_background_growth}), propagated them into the nonlinear power spectrum
and the effective-kernel bispectrum (Secs.~\ref{sec:bispectrum}), and combined the two statistics at the Fisher level
under Planck priors.

Adding the bispectrum improves the two-parameter figure of merit by factors
of $1.74$ (CPL), $1.64$ ($f(R)$) and $1.12$ (IDE). The gains are largest where
the new physics acts on nonlinear, mode-coupled scales e.g., the scale-dependent
growth of $f(R)$ and the growth history of CPL and smallest for IDE, whose
coupling enters our treatment only through a growth rescaling. The three-point
statistic rotates the parameter degeneracies inherited from the power spectrum,
which is the origin of the tighter joint constraints in
Fig.~\ref{fig:contours}.

These forecasts are exploratory in two specific senses: the covariances are
Gaussian, and the mode-coupling kernel is not recalibrated for the modified
dynamics. Both are the natural next steps includes an $N$-body calibration of the
beyond-$\Lambda$CDM bispectrum and the inclusion of non-Gaussian and super-sample covariance and both would sharpen, rather than overturn, the central result that the weak-lensing bispectrum is a useful and largely
independent lever on the dark sector.

\section*{Acknowledgements}

LFR and CBV acknowledge the National Research Foundation (NRF), South Africa for supporting their postdoctoral research through grant funding, and the National Institute for Theoretical and Computational Sciences (NITheCS) for additional financial support.

%\clearpage
\appendix

\clearpage
\bibliography{references}

\end{document}